\documentclass[journal]{IEEEtran}
\ifCLASSINFOpdf
\else
\fi

\usepackage[T1]{fontenc}
\usepackage[utf8]{inputenc}
\usepackage{lmodern}

\usepackage{adjustbox}

\begin{document}
%
\title{A systematic literature review on the development and use of mobile learning (web) apps by early adopters}
%
%
%

\author{Antonio~Ruiz-Martínez,
        Linda~Castañeda,
        and~Jesualdo T.~Fernández Breis 
\thanks{A. Ruiz-Martínez was with the Department
of Information and Communications Engineering, University of Murcia, Murcia,
Spain e-mail: arm@um.es.}
\thanks{L. Castañeda and J. T. Fernández Breis are with University of Murcia.}
}

\maketitle

\begin{abstract}
Surveys in mobile learning developed so far have analysed in a global way the effects on the usage of mobile devices by means of general apps or apps already developed. However, more and more teachers are developing their own apps to address issues not covered by existing m-learning apps. In this article, by means of a systematic literature review that covers 62 publications placed in the hype of teacher-created m-learning apps (between 2012 and 2017, the early adopters) and the usage of 71 apps, we have analysed the use of specific m-learning apps. Our results show that apps have been used both out of the classroom to develop autonomous learning or field trips, and in the classroom, mainly, for collaborative activities. The experiences analysed only develop low level outcomes and the results obtained are positive improving learning, learning performance, and attitude. As a conclusion of this study is that the results obtained with specific developed apps are quite similar to previous general surveys and that the development of long-term experiences are required to determine the real effect of instructional designs based on mobile devices. These designs should also be oriented to evaluate high level skills and take advantage of mobile features of mobile devices to develop learning activities that be made anytime at anyplace and taking into account context and realistic situations. Furthermore, it is considered relevant the study of the role of educational mobile development frameworks in facilitating teachers the development of m-learning apps.
\end{abstract}


%
\IEEEpeerreviewmaketitle

\section{Introduction}
\label{sec:introduction}

The evolution of mobile phones/smartphones has been vertiginous. We have seen that from 2007 to nowadays they have deeply improved their features such as computational processing, network speed, battery life and storage. Currently, we are able to do almost the same tasks we can do in a computer. These features with their quick adoption have made that these devices have been introduced for its use in a lot of fields. One of these fields that have benefit from them is education. This has lead to the birth of mobile learning (m-learning), which takes advantage of mobile devices to perform learning-teaching processes anytime, anywhere. Furthermore, this field is going to take more relevance due to COVID-19 since, as pointed out by Naciri et al.~\cite{Naciri_Baba_Achbani_Kharbach_2020}, m-learning in an unavoidable alternative in this scenario.

We have some studies that show that more and more m-learning is forming part for the teaching-learning process both formal and informal learning~\cite{Zydney_Warner_2016,Wu_Jim_Wu_Chen_Kao_Lin_Huang_2012}. Among its advantages we can point out teaching every time and every place, better connection with learners and offering a personalized and customized learning for the learner~\cite{Zydney_Warner_2016}. These advantages have attracted educators and researchers to apply it in learning scenarios, showing an important research between 2008 and 2016~\cite{Lai_2019}, which could be considered the main period for early adopters.

Many of the experiences developed by teachers are based on mobile applications (apps) that are already available for these devices. Thus, many of these apps are general-purpose apps for any smartphone such as Facebook, Dropbox, WhatsApp, Telegram, etc~\cite{Cetinkaya_Sutcu_2018,Aghajani_Adloo_2018,Robles_Guerrero_Llinas_Montero_2019,Smutny_Schreiberova_2020} and others are based on apps provided by the Mobile Operating System (MOS)~\cite{Jahnke_Kumar_2014,Mouza_Barrett-Greenly_2015,Engin_Donanci_2015,Ujakpa_Heukelman_Lazarus_Neiss_Rukanda_2018,Saritepeci_Duran_Ermis_2019}. However, we have already seen that a number of apps have been designed specifically for educational purposes~\cite{Zydney_Warner_2016,Sung_Chang_Liu_2016,Bano_Zowghi_Kearney_Schuck_Aubusson_2018}. This learning-oriented apps have shown better achievement that general ones~\cite{Sung_Chang_Liu_2016}. To this specific-purpose apps we name them as \textit{mobile learning apps} if they are developed for a specific MOS. If they are developed as a web application to be used in any mobile device we name them as \textit{mobile learning web apps}. To refer to these both types of apps that can be used in a mobile device we use the term \textit{mobile learning (web) apps} or (\textit{MLW apps} for short). 

These tools can be provided by a third party or being self-developed, although, in general, they are self-developed~\cite{Bano_Zowghi_Kearney_Schuck_Aubusson_2018}. The development of these self-developed apps requires, in general, that is made by a computer scientist although we have observed that some tools have been launched to facilitate that anyone without high expertise in programming apps can develop them~\cite{Ortega_2016,vedils_2018,AppInventorUM}. However, so far there is no much information on why someone decides to create a new app, with which methodologies is going to use it, which features they incorporate, etc. We do not have either much information if the features that are incorporated could be part of a framework for the development of apps without having expertise in it, which would facilitate the development of these apps for a wide community.

The purpose of our systematic literature review (SLR) is to analyse the different MLW apps that were developed by early adopters, which identify the need to create custom MLW apps, to see the different features they incorporated, which teaching methodologies were used and what results obtained. We also want to know whether these apps were developed by means of some kind of framework that facilitated its development and based on the analysis of the apps created which features should be supported. This analysis will allow us to know the motivations to develop this kind of applications and the kind of features that a m-learning app framework should provide to teachers to make easy the development of these tools. Finally, we would like to compare whether there are differences between the results obtained in general-purpose apps and these MLW apps. To the best of our knowledge, there is no survey that have analysed this issue so far.

Namely, the SLR that we have developed has the purpose of answering the following Research Questions (RQ) as for m-learning and the use of MLW apps:

\begin{itemize}
	\item RQ1. In which educational context were used the mobile (web) learning apps?
	\item RQ2. What was the purpose of the mobile learning (web) apps developed?
	\item RQ3. What were the results obtained in the experiences with the mobile learning (web) apps (positive, negative, mixed)?
	\item RQ4. What were the learning improvements (if there) that we were achieved with the use of specific mobile (web) applications and with were related to?
	\item RQ5. What pedagogical approaches were followed in the experiences with the apps?
	\item RQ6. What were the main features that the mobile learning (web) apps developed?
	\item RQ7. What were the development environments used to create these apps?
\end{itemize}

The paper is structured as follows. Section~\ref{sec:relatedwork} presents related work focusing on surveys about m-learning. Then, Section~\ref{sec:relatedwork} describes the procedure followed to develop this systematic literature review and, in Section~\ref{sec:results}, we present the results obtained. Next, we answer all the research questions and discuss the results in Section~\ref{sec:discussion}. Finally, in Section~\ref{sec:conclusions} we present the conclusions obtained.

\section{Related Work}
\label{sec:relatedwork}

As m-learning is a subject with an important interest in the research field, there are a lot of publications and this has derived in a plethora of surveys with the aim of providing a comprehensive view of the research made regarding m-learning. In Table~\ref{tab:PreviousSurveys}, we show the main surveys found in the literature that cover the apps developed by early adopters and we indicate the main information that characterize them: the period covered by the survey, the subject in which they are focused on, the educational levels the survey covers, the number of papers analysed and, finally, the main issue reviewed.

 
\begin{table*}[!htb]
	\centering
		\caption{Summary of previous reviews}
	\begin{adjustbox}{width=\textwidth}

		\begin{tabular}{|c|c|c|c|c|c|}

\hline

\textbf{Reference} & \textbf{Timeline} & \textbf{Subject} & \textbf{Educational Level} & \textbf{\#papers} & \textbf{Issue reviewed}  \\ \hline

\cite{Wu_Jim_Wu_Chen_Kao_Lin_Huang_2012} & 2003-2010 & Any & Any & 164 & Trends from mobile learning \\ \hline
\cite{Liu_Scordino_Geurtz_Navarrete_Ko_Lim_2014} & 2007-September 2012 & Any & Any & 63 & Study of m-learning in K-12 \\ \hline

\cite{Al_Zahrani_Kumar_2015} & 2009-2013 & Any & Higher education & 8 & Meta-analysis about experiences of m-learning and its acceptance and use \\ \hline
\cite{Alrasheedi_Capretz_Raza_2015} & 2005-2013 & Any & Any & 30 & Critical factors for success of m-learning in Higher Education\\ \hline
\cite{Sung_Chang_Liu_2016} & 1993-2013 & Any & Any & 110 & Effect of integrating mobile devices in teaching and learning on learning performance\\ \hline

\cite{Baran_2014} & -August 2014 & Any & Teachers' education & 37 & Mobile learning and teacher education\\ \hline
\cite{Zydney_Warner_2016} & 2007-2014 & Science learning & School & 37 & Mobile apps for science learning\\ \hline

\cite{Pena-Ayala_Cardenas_2016} & 2010-1st quarter 2015 & 	Any & Any & 105 & Mobile, ubiquitous and pervasive learning study \\ \hline
\cite{Chee_Yahaya_Ibrahim_Hasan_2017} & 2010-2015 & Any & Any & 144 & Trends in mobile learning\\ \hline
\cite{Crompton_Burke_Gregory_2017} & 2010-2015 & Any & PK-12 & 113 & Study of m-learning in PK-12 \\ \hline

\cite{Bano_Zowghi_Kearney_Schuck_Aubusson_2018} & 2003-2016 & Math \& Science & Secondary school & 60 & High quality empirical evidence on m-learning in secondary school science and maths education \\ \hline
\cite{Crompton_Burke_2018} & 2010-2016 & Any & Higher education & 72 & Mobile learning research in higher education settings\\ \hline
\cite{Hwang_Fu_2019} & 2007-2016 & Language & Any & 93 & Design and application of mobile language learning \\ \hline

\cite{chee2017review} & NS & Any & Any & NS & General literature review \\ \hline
\cite{Lai_2019} & 2000-2016 & Any & Any & 100 & Trends of m-learning from 100 highly cited papers \\ \hline
\cite{fu2018trends} & 2007-2016 & Any & Any & 90 & Mobile technology-supported collaborative learning\\ \hline
\cite{Nikou_Economides_2018} & January 2009-February 2018 & Any & Any & 43 & Survey of mobile assessment literature\\ \hline
\cite{Chung_Hwang_Lai_2019} & 2010-2016 & Any & Any & 63 & Insights and trends of mobile learning based on experimental mobile learning studies\\ \hline

		\end{tabular}
	\end{adjustbox}
	\label{tab:PreviousSurveys}
\end{table*}

We can see that the interest in summing up the information regarding m-learning comes from 2012, aiming at covering the very initial experiences performed with these recent devices. We can also see that the interest in reviewing this literature has been made in surveys mainly published in 2018 and 2019 and covering mainly until 2016. We can also point out that most of the surveys cover any subject, although some of them are focused on Science, Maths and Language. As for the education level covered, most of them cover all levels but there are also some of them focused on Higher Education or schools. As for the number of studies analysed, we can see that the first survey is the one which has reviewed more papers and that more recent surveys cover a more reduced number of papers due to they are focused on particular issues regarding m-learning or because they have decided to fence in the study to the most relevant papers since currently the number of papers in the m-learning field is huge. 

As for the aspects reviewed we can see that they cover different issues such as trends, its usage in different educational levels or for different purposes such as assessment or collaboration. However, to the best of our knowledge, none of the surveys developed so far has focused on analysing the experiences developed with the use of specifically developed MLW apps. 

%
%
%
%

\section{Methods}
\label{sec:methods}

The research method we have used to perform the review of the literature regarding the development and use of mobile learning (web) apps is a Systematic Literature Review (SLR). For this SLR, we have followed the guides proposed by \cite{Kitchenham_Charters_2007}. We decided to use a SLR because it follows a rigorous methodology that could be reproduced by other researchers unlike traditional reviews.

In this section, we describe the different steps we have followed to carried out the SLR and manage to answer the different research questions we have exposed previously.

%
%

\subsection{Search strategy}

We decided to use as search databases: Scopus, EBSCO, and Google Scholar. We use them because we consider them the most suitable for this field. 

As keywords we considered: m-learning, app, authoring tool, development, teaching, methodology, and activity. Based on these keywords as search string we used: (“m-learning” OR “mobile learning”) AND ((app OR application) OR “authoring tool” OR “authoring tools”) AND teaching AND (methodology OR method OR design OR activity) AND (development OR creation OR design). 

We launched our queries to the databases in the period from 2012 to 2017. In the case of Google, we limited the query to 100 results since we wanted to create this SLR based only on the most relevant papers.

\begin{figure}[htbp]
	\centering
		\includegraphics[scale=0.25]{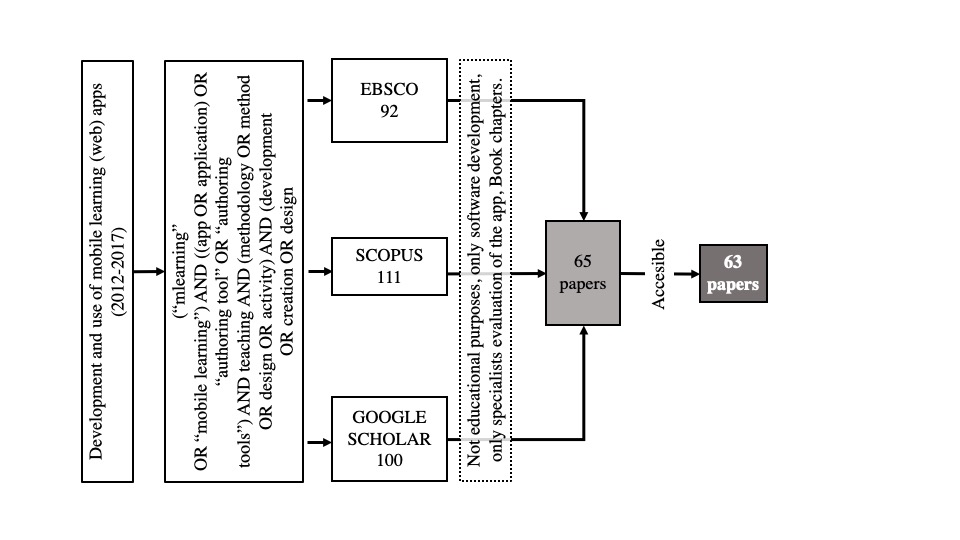}
	\caption{SLR process and publications analysed}
	\label{fig:FigureNumberofPublications}
\end{figure}

Our initial query in the databases obtained 303 papers distributed (see Figure~\ref{fig:FigureNumberofPublications}). After removing repeated papers, the result became 244 papers. Next, we enumerate the inclusion and exclusion criteria we have applied.


\subsection{Inclusion and exclusion criteria}

The inclusion criteria are:
\begin{itemize}
	\item Must be published between 2012 and 2017.
	\item Must be a journal article or a conference paper.
	\item The focus must be on mobile learning.
	\item Must include data about the use of the specific (web) app by the students or the teacher or any evaluation regarding its use or performance. Only software description will not be considered.
	\item Must include educational activities.
	\item Must include the use of mobile devices.
	\item Must include the use of mobile (web) applications.
\end{itemize}

The exclusion criteria are:
\begin{itemize}
	\item Mobile learning not used for educational purposes.
	\item Papers only presenting developments oriented to create m-learning management systems or MLW apps.
	\item Papers that only evaluate usability of m-learning apps by a specialist in the field.
	\item Book chapters.
\end{itemize}

The number of papers after applying inclusion and exclusion criteria was 65. From the selected articles, we were not able to access to the text of 3 of them. Therefore, the final number of papers to be analysed is 62. 

\begin{table*}
	\centering
		\caption{Papers selected}
	\begin{adjustbox}{width=\textwidth}
	
		\begin{tabular}{|c|c|c|c|c|c|}
			\hline
			\textbf{Source ID} & \textbf{Authors} & \textbf{Learner Level} & \textbf{Domain} & \textbf{Duration} & \textbf{Number of students} \\
			\hline
			S01 & \cite{Abd_Husain_2014} 															& Primary school & Science & ST & NS\\
			S02 & \cite{Adam_Kioutsiouki_Karakostas_Demetriadis_2014}  & University &	Engineering &	ST & NS\\
			S03 & \cite{Ahmad_Shahid_2015} 														& School (8-12) & Education & MT & NS\\
			S04 & \cite{Astra_Nasbey_Nugraha_2015}											& High school & Science & NS & NS\\
			S05 & \cite{Bogdanovic_Barac_Jovanic_Popovic_Radenkovic_2014} & University & Business & LT & NS \\
			S06 & \cite{Boticki_Baksa_Seow_Looi_2015} 									& Primary school & Science & LT & 305 \\
			S07 & \cite{Boticki_Barisic_Martin_Drljevic_2013} 					& University & Computer Science & LT & 120\\
			S08 & \cite{Boticki_Wong_Looi_2013} 												& Primary School & Maths \& Language & ST & 28\\
			S09 & \cite{Browne_Anand_2013} 														& University & Computer Science & MT & 70-49 \\
			S10 & \cite{Cabielles-Hernandez_etal_2014} 								& Autism center & Education & ST & 2 \\
			S11 & \cite{Carroll_Kop_Thomas_Dunning_2015} 							& Nursery/primary school & Language & NS & 50\\
			S12 & \cite{Chang_Chatterjea_etal_2012}										& University & Geography & ST & 12 \\
			S13 & \cite{Chua_Balkunje_2012} 														& University & Several domains & LT & 55 \\
			S14 & \cite{Cocciolo_Rabina_2013} 													& University (post) & History & ST & 31\\
			S15 & \cite{Correa_etal_2013} 															& Elementary school & Geometry & NS & 5\\
			S16 & \cite{Dirin_Nieminen_2015}														& University \& Drive license school & Computer Science, driving license \& business & ST & 12\\
			S17 & \cite{Fernandez-Lopez_etal_2013} 										& Special and elementary school & Special Education & LT & 39 \\
			S18 & \cite{He_Ren_Zhu_Cai_Chen_2014} 											& Kindergarten & Education & ST & 40 \\
			S19 & \cite{Herrera_Sanz_2014} 														& University (post) & Engineering & ST & 13\\
			S20 & \cite{Huang_Yang_Chiang_Su_2016} 										& Elementary school & English & ST & 80\\
			S21 & \cite{Jeno_Grytnes_Vandvik_2017} 										& University & Biology & ST & 71\\
			S22 & \cite{Jou_Lin_Tsai_2016} 														& University & Engineering & LT & 87\\
			S23 & \cite{Kamaruzaman_Zainol_2012} 											& Secondary school & English & ST & 350\\
			S24 & \cite{Kim_Kim_Han_2013} 															& Teacher training college & Information Communication Technologies & ST & 34\\
			S25 & \cite{Kwang_Lee_2012} 																& University & NS & ST & 58 \\
			S26 & \cite{Lai_Hwang_2015} 																& Elementary school & Art design & ST & 103\\
			S27 & \cite{Land_Zimmerman_2015} 													& Any level & Botanic & NS & 53\\
			S28 & \cite{Lehmann_Sollner_2014} 													& University & Computer Science & ST & 85\\
			S29 & \cite{Lu_Meng_Tam_2014} 															& Primary school & Language & MT & 17\\
			S30 & \cite{Martin_Ertzberger_2013} 												& University & Education & ST & 109 \\
			S31 & \cite{Melero_Hernandez-Leo_Manatunga_2015} 					& Secondary education & Art & NS & 82\\
			S32 & \cite{Melero_Hernandez-Leo_Sun_Santos_Blat_2015} 		& Secondary school & Maths, natural science, geography, and art history & ST & 81\\
			S33 & \cite{Mintz_2013} 																		& Special school & Education & LT & 26\\
			S34 & \cite{Mokgonyane_etal_2017} 													& Primary school & Maths & ST & 20\\
			S35 & \cite{Muthukumarasamy_A_2013} 												& University & Computer science, physics, chemistry \& maths & ST & 60\\
			S36 & \cite{Noguera_etal_2013} 														& University & Physiotherarpy & ST & 60\\
			S37 & \cite{Ortiz_etal_2015} 															& University & Engineering \& Telecommunication & NS & 163\\
			S38 & \cite{Palomo-Duarte_Berns_Dodero_Cejas_2014} 				& University & Foreign language & ST & 120\\
			S39 & \cite{Pereira_2016} 																	& University & Computer science & LT & 23\\	
			S40 & \cite{Pham_Chen_Nguyen_Hwang_2016} 									& Any level & English & LT & 2774 \\
			S41 & \cite{Popovic_Markovic_Popovic_2016} 								& University & Computer science & LT & 128\\
			S42 & \cite{Reis_Escudeiro_Escudeiro_2012} 								& University & Graphics systems and multimedia & NS & NS\\
			S43 & \cite{Rensing_Tittel_Steinmetz_2012} 								& University & Engineering & NS & NS \\
			S44 & \cite{Rubegni_Landoni_2014} 													& Primary school & Language & ST & 43\\
			S45 & \cite{Santos_etal_2014} 															& High school and university & Art history and botany & ST & 122\\
			S46 & \cite{Schmitz_Klemke_Walhout_Specht_2015} 						& School (12-18) & Healt & LT & 238\\
			S47 & \cite{Shu-Chun_Sheng-Wen_Pei-Chen_Cheng-Ming_2017} 	& Different students (18-30) & English & ST & 90\\
			S48 & \cite{Skiada_Soroniati_Gardeli_Zissis_2014} 					& Dyslexia school (8-12) & Dyslexia & LT & 5\\
			S49 & \cite{Su_Cheng_2015} 																& Primary school & Botanic & MT & 102\\
			S50 & \cite{Sun_Chang_2016} 																& University & Botanic & NS & 64\\
			S51 & \cite{Sung_Hwang_Liu_Chiu_2014} 											& University & Architecture & ST & 56\\
			S52 & \cite{Toktarova_Blagova_Filatova_Kuzmin_2015} 				& University & Engineering & LT & 86\\
			S53 & \cite{Vasquez-Ramirez_etal_2014} 										& High school & Chemistry & NS & 22\\
			S54 & \cite{Wald_Li_Draffan_2014} 													& University & NS & NS & 10\\
			S55 & \cite{Waldmann_Weckbecker_2013} 											& University & Family Medicine Course & ST & 14\\
			S56 & \cite{Wang_2013} 																		& University & Ecology & NS & 167\\
			S57 & \cite{Wang_2016} 																		& High school & Language & ST & 56\\
			S58 & \cite{Wen_Zhang_2015} 																& University & Medicine & LT & 180\\
			S59 & \cite{Wu_2015} 																			& University & Language & MT & 70\\
			S60 & \cite{Zbick_Nake_Jansen_Milrad_2014} 								& High school & Science & NS & NS\\
			S61 & \cite{Zbick_Nake_Milrad_Jansen_2015} 								& High school & Sport, biology, geography, \& language  & ST & 13T\\
			S62 & \cite{Zhao_Wu_Chen_2017} 														& University & Electronic engineering & ST & 20\\
			\hline
		\end{tabular}
		
		\end{adjustbox}
	\label{tab:PapersSelected}
\end{table*}

\subsection{Data collection}

Based on the information we want to analyze from the research questions, we decided to extract the following information from each paper:

\begin{itemize}
	\item The source and full reference. We annotate the search engine(s) where the manuscript was located and a full reference for its citation.
	\item Number of cites received (we have stored the maximum number among the different sources).
	\item Article type (conference/journal). We identify whether the manuscript was published in a conference or in a journal and its name.
	\item Authors. We have considered the number of authors that have participated in the paper.
	\item Authors' background.  We have analysed author's profile to determine their background and whether the author is a computer engineer or not.
	\item Research methods. The research method used in the paper has been gathered and we indicate whether it is quantitative, qualitative or is a mixed method.
	\item Research purpose. Based on~\cite{Crompton_Burke_2018}, we have identified as possible research purposes: student achievement, student perception, pedagogy, factors influencing, device/app (investigating a specific application or mobile learning system). Additionally, we have added: teacher perception, whether the research was interested in obtaining information about teachers' perception of the use of the mobile app, and supporting teachers, whether the app was designed to support teachers' activities.
	\item Theoretical foundations. Following the~\cite{Hannafin_Hannafin_Land_Oliver_1997}'s classification regarding how the theoretical foundations have been applied to the development of the app we classify as: grounded, cited, and not provided.
	\item Learner level. The students' academic level where the experience has been developed, e.g., elementary, secondary, etc.
	\item Number of students. We have collected the number of users that have participated in the experience. 
	\item Type of mobile devices used. We gathered information both on the type of mobile device and MOS used.
	\item Fields of Education and Training classification. This item stores information on students' education/training field where the experience has been conducted according to International Standard Classification of Education: Fields of Education and Training 2013 (ISCED-F 2013)~\cite{ISCED_2013}.
	\item Learning settings. With reference to~\cite{Hwang_Tsai_Yang_2008} and~\cite{Crompton_Burke_2018}, we identified the settings where the learning experience was developed: classroom, city, campus, environment, event, museum, landscape, or at home. In some cases, it can also marked as not specified if the authors did not specify the learning setting. We used this classification as long as it was clear the setting, if the app can be used in every place we assigned the key \textit{outdoors} and we assigned \textit{school test} when the experience has been only developed for a test.
	\item Educational contexts. The context used in the studies follow the classification presented by~\cite{Bano_Zowghi_Kearney_Schuck_Aubusson_2018}, where the authors classify them into formal learning, semi-formal learning and informal learning. 
	\item Learning activities. We classified the apps according the type of learning activities they are supposed designed to be done with during the teaching-learning experience. We have followed~\cite{Bloom_1956}'s classification.
	\item Learning impact. The different experiences have been designed to provoke a learning impact that could be classified as~\cite{Domingo_Gargante_2016}: providing new ways to learn, increasing engagement to learning, fomenting autonomous learning, facilitating access to information, and promoting collaborative learning.
	\item Learning outcome. We collect whether the results of experience have been positive, negative, or mixed.
	\item Research question information. For each paper, we have included all the information needed to answer each research question presented in the introduction.
\end{itemize}

Table~\ref{tab:PapersSelected} depicts the data for each survey paper selected. The rest of information is shown in Table~\ref{tab:NumberPublicationsYear}.



\subsection{Data analysis}

After performing the search, applying inclusion and exclusion criteria and collected all the information specified in the previous section, we have started the analysis phase where all the data collected has been analysed to provide statistical information on the different information as well as to answer all the research questions defined for our SLR. The results are presented in Section~\ref{sec:results} and the answer to the different research questions is discussed in Section~\ref{sec:discussion}.

\section{Results}
\label{sec:results}

In this section, we provide the data collected from the papers analysed regarding its features and its publication (such as the number of publications per year, publication type, research methods), features of the apps developed for the experiences and of the learning experiences.

\subsection{Publications per year}

In Table~\ref{tab:NumberPublicationsYear} appears the number of papers selected according to the year the article was published. The majority of the papers analysed were published after 2014.

\begin{table}
	\centering
	\caption{Results classified by publication year}
		\begin{tabular}{|c|c|}

\hline
\textbf{Year} & \textbf{Number of publications}  \\ \hline
2017 & 4 \\ \hline
2016 &	7\\ \hline
2015 &	16\\ \hline
2014 &	15 \\ \hline
2013 & 14\\ \hline
2012 & 6 \\ \hline

		\end{tabular}
	
	\label{tab:NumberPublicationsYear}
\end{table}

\subsection{Type of publications}

From these 62 publications, 44 papers were published in journals and 18 in conferences. These papers have been published in 43 different kind of publications: 30 different journals and 13 different conferences. 

The publications from which we have selected more papers are: Computers \& Education, with 11 articles; International Conference on Advanced Learning Technologies, with 3 papers; Procedia Computer Science, with 2 articles; Journal of Educational Technology \& Society, with 2 articles, Journal of Computer Assisted Learning, with 2 articles; International Conference on Technological Ecosystems for Enhancing Multiculturality, with 2 papers; Eurasia Journal of Mathematics, Science \& Technology Education, with 2 articles; Computer Applications in Engineering Education, with 2 articles; and, finally, British Journal of Educational Technology, with 2 articles. The rest of publications only has an article or paper.

\subsection{Research methods}
The research made in the selected papers has followed one of the following research methods: quantitative, qualitative or mixed methods. In Table~\ref{tab:ResearchMethods}, we show the results obtained for the different methods and the number of times they have been used. As can be seen, we can point that the most used ones are the quantitative and qualitative methods (they cover 87.1\%). Mixed research methods are hardly used.

\begin{table}
	\caption{Research Methods}
	\centering
		\begin{tabular}{|c|c|}

\hline
\textbf{Method} & \textbf{Number of times used}  \\ \hline
Quantitative & 28 \\ \hline
Qualitative & 26 \\ \hline
Mixed Methods & 8 \\ \hline
		\end{tabular}
	\label{tab:ResearchMethods}
\end{table}


\subsection{Theoretical foundations}

We have also analyzed whether in the research presented in the paper, the authors present the theoretical foundations in which they have based on. Our analysis show that 39 papers (62.93 \%) do present the theoretical foundations in which they have based on. 

We can point out a number of foundations in the experiences: field-based learning, inquiry, collaborative work/learning, self-directed learning, badges for learning, teacher-directed, game-based learning, placed-based learning, situated learning, problem-based learning, motivation and engagement foundations, peer-assessment, context-sensitive learning, guided participation, here and now learning, informal learning, ubiquitous learning, Taxler's designing of learning model, constructivist theory, Engeström’s expansive activity model, social constructivist approach, annotation strategy, inquiry-based learning, flow learning, flipped classroom, motivation theory, self-directed learning, microlecture, m-Learning, and ubiquitous learning, Mobile-assisted language learning (MALL) and authentic learning.

\subsection{Research purpose}
In the analysis of the research purposes, we have based on the classification introduced by~\cite{Crompton_Burke_2018}. They classify them into: student achievement, student perception, pedagogy, factors influencing and device/app. Furthermore, we have also included a new category: teacher support, since some tools are not for students but for teachers. In general, in most of the papers, there is more than a research purpose. In Table~\ref{tab:ResearchPurposes} we can see the results of the analyzed papers. The three main issues are: device/app, student perception, and student achievement.

\begin{table}
	\caption{Research Purposes}
	\centering
		\begin{tabular}{|c|c|}

\hline
\textbf{Research purpose} & \textbf{Number of occurrences}  \\ \hline
Student achievement & 34 \\ \hline
Student perception & 53 \\ \hline
Pedagogy & 12 \\ \hline
Factors influencing & 8 \\ \hline
Device/app & 60 \\ \hline
Teacher support & 1 \\ \hline

		\end{tabular}
	\label{tab:ResearchPurposes}
\end{table}

Next, we analyse in detail the apps following the classification used by~\cite{Domingo_Gargante_2016} where they classify the m-learning impact that experiences can cause in those that are for introducing new ways of learning, increasing the engagement, fomenting autonomous learning, facilitating access to the information, and promoting collaborative learning. 

\begin{figure}[tbp]
	\centering
		\includegraphics[scale=0.5]{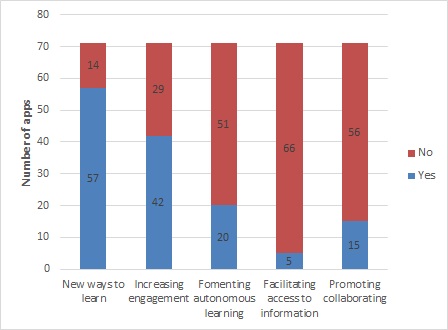}
		\caption{Mobile apps learning impact}
	\label{fig:FigureOutcomesEvaluated}
\end{figure}

In Figure~\ref{fig:FigureOutcomesEvaluated} we show the different apps classified according to the m-learning impact they were designed to. We can observe that three main objectives of the apps were a new way to learn, followed by increasing engagement and, finally, promoting autonomous learning.

\subsection{Citations}

From the selection, the paper which has received more citations (351, according to Google Scholar results) is from~\cite{Martin_Ertzberger_2013} where the authors introduce a new kind of m-learning which they named as \textit{Here and Now} m-learning showing positive results as for achievement and attitude. In average, the papers received 32 cites. The mode is 4.

\subsection{Apps features}
\label{sec:appsfeatures}

In this section, we analyze the different features that characterize the apps involved in the different papers from different point of views: we have considered its development process and whether they were developed by a computer programmer, by a professional of educational technology or in collaboration between both. We have also considered the kind of devices and operating system used, what tools were used for its development of the app, the type of apps developed according their purpose and, finally, their functional features.

\subsubsection{Background developers' apps}

In each publication we have analyzed whether the app design was based on (collaboration with) professionals of the educational technology field and whether they were developed by no computer scientist. In some papers the authors present the results of more than one app. Thus, the total number of apps described in the 62 publications is 71.

Our analysis shows that in 35 apps (49,3\%), educational technology experts participated in the design. We can also see that there are only 10 applications that seem that were not developed by researchers/educators with background in computer science (CS). The rest (36) were developed by researchers/educators with background in CS.

We can also mention that 26 papers (41,93\%) are from authors that only come from the technological field, i.e., the application was designed and developed by a computer science engineer. Therefore, in most papers (58,07\%) there was some collaboration among computer science engineers and educational technologists.

\subsubsection{Device/OS for the apps}
The apps were designed for different kind of configurations regarding the device to use and/or Mobile Operating System (MOS). There are apps that also were developed for different devices/MOSs. Thus, there are apps that were developed for Android devices or even for more specific configurations taking into account the type of device, for example, apps that were designed for Tablets Android. The number of combinations we have found are 11. We have also found experiences were there was no indication about app and/or MOS. 

From the analysis, we can point out that there are 11 apps (15,49 \%) that were designed to be executed in any kind of device (because they were based on Web or they had different version of the app for the different MOS). However, the most popular configuration is the use of Android for different kind of devices such as mobile phones or tablets. Namely, the analysis of the apps shows that there are 25 apps that can be used in any kind of Android device (35,21 \%). Finally, we can find some apps developed for iPads, iPods, iOS or Windows Phone.

\subsubsection{Apps development tools}
\label{sec:appsdevelopmenttools}
Most of them (53 apps) were exclusively developed with tools that generate native apps for a specific MOS. As Mobile Web apps, we can find 14 apps which were exclusively created as mobile web apps. Finally, we have 9 apps that were developed both as mobile apps and MLW apps.

%
For app development, different tools were used. In general, we can find that developers used the tools provided by the mobile platform or web tools for the development but we have also seen that there are apps that were developed using frameworks or platforms that support building apps using high level tools (what we call an app framework/platform) and no developer-specific tools. 

In some papers the authors present the framework/platform and the apps developed based on that framework. In this case, we can mention we found 9 apps that are presented as a proof of concept of the framework that is introduced with the app. The frameworks presented are: mLUX~\cite{Dirin_Nieminen_2015}, mLearn4web~\cite{Zbick_Nake_Milrad_Jansen_2015,Zbick_Nake_Jansen_Milrad_2014}, AthenaTV~\cite{Vasquez-Ramirez_etal_2014}, mCSCL - content-independent collaborative mobile learning~\cite{Boticki_Wong_Looi_2013}, MCL Application~\cite{Kwang_Lee_2012}, and docendo~\cite{Rensing_Tittel_Steinmetz_2012}. 


We aslo have apps that indicate that they were developed based on an existing framework. The frameworks used for these apps are: AppInventor used in~\cite{Adam_Kioutsiouki_Karakostas_Demetriadis_2014}, Lectora Inspire in~\cite{Martin_Ertzberger_2013}, ARLearn in~\cite{Schmitz_Klemke_Walhout_Specht_2015}, and Basic4Android in~\cite{Wu_2015}. 

Finally, we have a set of platforms that allow the teacher to create mobile learning experiences from the same type (e.g., field trips or educational games). Within this category we have SamEx, MobiTop, GeoStoryTeller, Picaa, IM2Learn, QuesTInSitu: The Game, HANDS, mLMS, "@dawebot", PACARD, mTester, Mobile Gamification Learning System, Prompt-based annotation mobile learning system, Programming Mobile Applications for Android, AthenaTV, Synote, JMNS, Cross-device Mobile-Assisted Classical Chinese (CMACC) system, MMLS, and Android-based mobile educational platform (AEPS). This is the most numerous group.

\subsubsection{Type of apps}
\label{sec:typeofapps}

\cite{Domingo_Gargante_2016} provide a generic classification of the apps depending of the purpose they could be used. In this classification, we have the following categories: Learning Skills Apps (LSA), Informational Management Apps (IMA), and Content Learning Apps (CLA).

\begin{figure}
	\centering
		\includegraphics[scale=0.4]{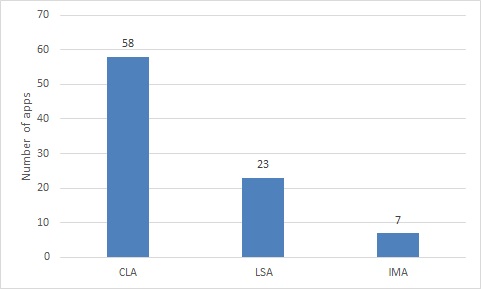}
	\caption{Classification of the apps based on purpose}
	\label{fig:TypeAppsPurpose}
\end{figure}

\begin{figure}
	\centering
		\includegraphics[scale=0.4]{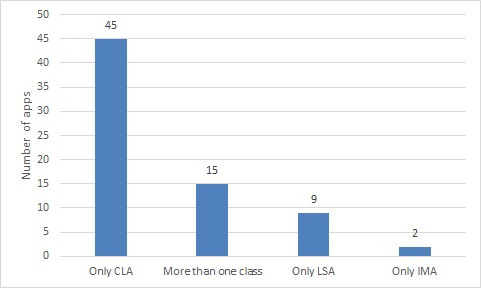}
	\caption{Classification of the apps based on purpose by single classes}
	\label{fig:TypeAppsPurposeSingleClasses}
\end{figure}

The analysis of the apps used in the different experiences show, in Figure~\ref{fig:TypeAppsPurpose}, that the category that registers more number of apps is CLA with 58 apps, followed by LSA with 23 and, finally, IMA with only 7 apps. There are 62 apps, therefore, there are apps that have been classified in more than one category. Namely, as can be seen in Figure~\ref{fig:TypeAppsPurposeSingleClasses}, there are 15 apps (21,12\%) that are within more than one category. Being the combination of CLA and SLA, the most frequent one. We have also found out that there are 45 apps (63.38\%) which are only CLA, 9 apps (12.68\%) that are only SLA, and 2 apps (2.81\%) that are only IMA.

\begin{figure}
	\centering
		\includegraphics[scale=0.4]{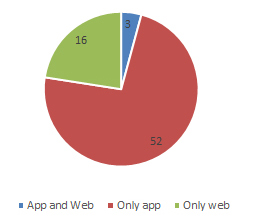}
	\caption{Number of apps depending on its development.}
	\label{fig:NumberApps}
\end{figure}

From the development point of view (see Figure~\ref{fig:NumberApps}), 52 (73\%) mobile applications have been developed as specific apps for a mobile operating system, 16 mobile applications were designed as mobile web applications and, finally, there are 3 (4\%) mobile apps that have been developed as both an app and a mobile web application.

We can also mention that 42 of these apps (59,15\%) require the use of a server for working and the rest - 29 (40,85\%) - can work as a standalone application. We can also mention that 4 apps (5,63\%) are integrated with a Learning Management System (LMS).


Finally, we can remark that 19 apps (26,76\%) were designed to be used in an outdoor environment.

\subsubsection{Functional features}

In this section, we present the analysis developed on the different functional features the apps have incorporated. 

\begin{figure*}
	\centering
		\includegraphics[scale=0.4]{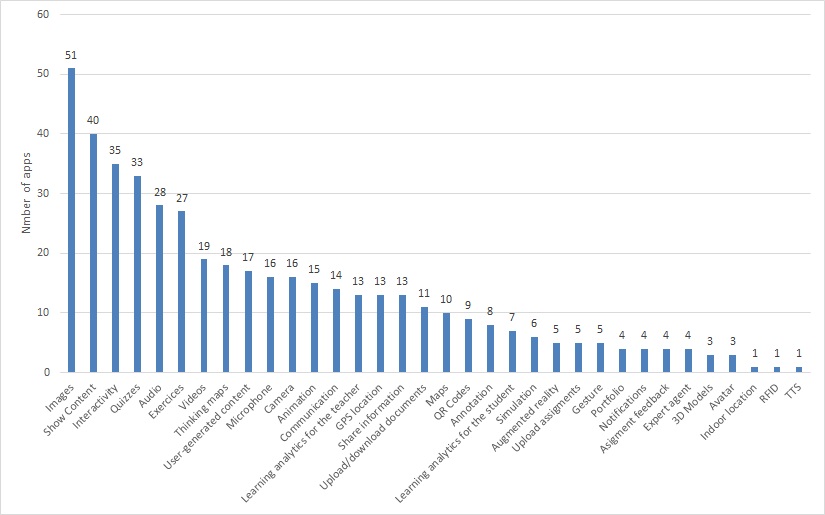}
	\caption{Apps Functional features.}
	\label{fig:AppsFunctions}
\end{figure*}

As might be expected the use of images and the incorporation of content is a key feature in most of the applications (see Figure~\ref{fig:AppsFunctions}). Almost in the same level of usage, we can also point out interactivity, the incorporation of quizzes, audio, and exercises. Then, there is a second level of features where videos, thinking maps, etc.

In Figure~\ref{fig:AppsFunctions} we can also see that there are 13 apps that provide learning analytics to the teachers and that 7 apps are providing analytics to the students. Learning analytics is becoming a trend and we can see that its usage associated to mobile devices can better support the learning process~\cite{Rohloff_et_al_2018,TABUENCA201553}.


Other interesting features that are being incorporated recently in the MLW apps are the use avatars, RFID and Text To Speech (TTS)~\cite{Dirin_Nieminen_2015,Pham_Chen_Nguyen_Hwang_2016,Arai2019}.







\subsection{Features of the learning experiences}

This section analyses the features of the learning experiences: their learner type, number of students who have participated, fields of education and training classification where the experiences were developed, their learning settings, the educational context where they were developed, the outcomes of their results, who the owner of the mobile devices is, whether the use of the app is compulsory in the experience, the measured outcomes, the teaching methodologies used, and the learning activities carried out.

\subsubsection{Learner Type}

\begin{figure}
	\centering
		\includegraphics[scale=0.5]{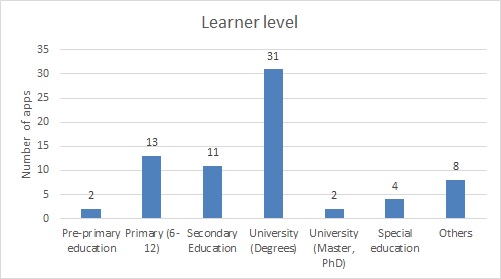}
	\caption{Number of apps used for each learner level}
	\label{fig:LearnerLevel}
\end{figure}

We analysed the educational level for which the apps developed were used. Figure~\ref{fig:LearnerLevel} shows that there are 2 apps used in experiences made with kindergarten students, 13 with students of primary education, 11 with secondary students, 31 with undergraduate students, 2 with graduated students, 4 with students in special education courses (e.g., autism, dyslexia) and, finally, we have 8 experiences that were made for teachers (2), future professionals (family medicine courses), business, driving license schools or field trips (2) or any level. Thus, we observed that the most target group come from experiences developed in the university (31).

\subsubsection{Number of students}
\begin{figure}
	\centering
		\includegraphics[scale=0.3]{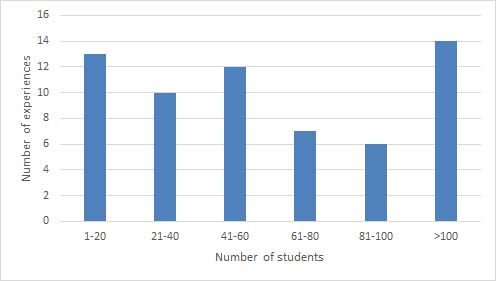}
	\caption{Number of students in the experiences}
	\label{fig:NumberStudents}
\end{figure}

The analysis of the experiences show a diverse participation of students in them. The experience with the major participation has been in the work of~\cite{Pham_Chen_Nguyen_Hwang_2016} with 2774. In average, the participation of students in the experiences is 115. The analysis depicted in Figure~\ref{fig:NumberStudents} indicate that the number of experiences with more than 100 students is the most numerous group. Then, we have experiences that have participated between 1 and 20 students and, then, the experiences with 41 and 60 students. We can also see that the number of experiences with 100 or more students is higher than the experiences than were performed with than less 100 students.

\subsubsection{Experience duration}

\begin{figure}
	\centering
		\includegraphics[scale=0.35]{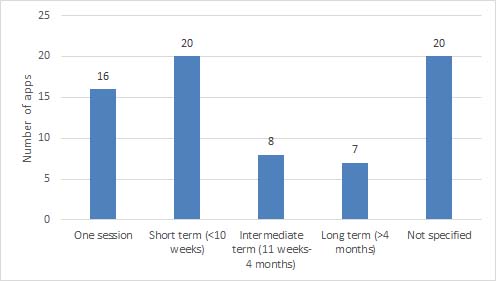}
	\caption{Duration of the different experiences with the apps}
	\label{fig:ExperienceDuration}
\end{figure}

In Figure~\ref{fig:ExperienceDuration} we can see that most of the experiences last one session or at most 10 weeks (short term experiences). The number of experiences that report long term periods is quite reduced (only 7). We can also point out that main of experiences with the apps have not specified the time they were used.

\subsubsection{Fields of Education and Training classification}
The apps developed were used in different disciplines. The disciplines with the number of experiences are indicated next: Science (7), Engineering (23), Architecture (1), Business (1), Maths (5), Humanities (17), Social Sciences (10), Economics (2), Experimental sciences (7), Health Science (4) and others (driving license courses or not specified). 

As can be seen, most of the experiences reported are within the field of Engineering. Within Engineering we can point out Computer Science with 15 experiences. Within Social Science, we can remark Education with 9.

\subsubsection{Learning setting}
\begin{figure}
	\centering
		\includegraphics[scale=0.5]{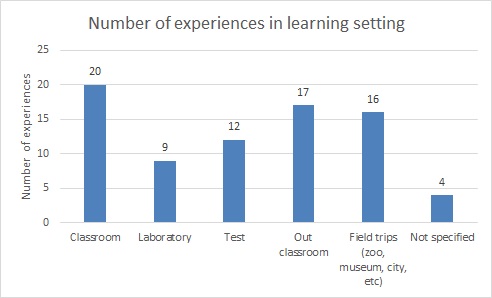}
	\caption{Number of experiences in learning setting}
	\label{fig:LearningSetting}
\end{figure}

From the analysis we have made on which learning setting the apps have been used (see Figure~\ref{fig:LearningSetting}), we can indicate that the different experiences presented have been used in 78 different settings. 

In the setting that more experiences were developed is in the classroom, with 20 (25,64\%). Next, we have found 17 experiences (21,79\%) which were developed outside the classroom (without requiring any special place). There are 12 experiences (15,38\%) which have only been developed as a test for the app (they are not reproduced in normal conditions of a classroom). Laboratories is a setting where 9 experiences (11,54\%) were developed. Next, we have a set of setting which represent the use of the app outside the classroom but in a specific scenario such as field trips (8 experiences - 10,26\%), museums (4 experiences - 5,13\%), and city environment (2 experiences - 2,56\%). Finally, there are 4 experiences (5,14\%) where the authors have not specified the learning setting.

We want to point out that the experiences made in a real-context or in authentic situations were 19 (30.64\%).

In the different experiences, 44 apps were used in the classroom or in an activity related to the classroom such as a planned outdoor activity or in blended learning. We have found that 30 apps are to be complementary to the lectures. From these apps, 10 are to be used in the classroom and out of the classroom.

\subsubsection{Educational context}
\begin{figure}
	\centering
		\includegraphics[scale=0.45]{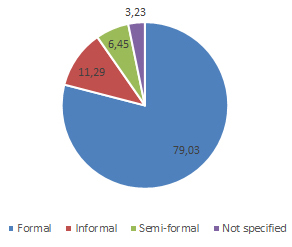}
	\caption{Experiences (\%) in function of Educational Context}
	\label{fig:EducationalContext}
\end{figure}

Analysing the educational context where the different experiences were developed, the results show (see Figure~\ref{fig:EducationalContext}) that most of them were developed in a formal context (49 experiences - 79,03\%). Then, we have found 7 experiences (11,29\%) developed in an informal context and 4 experiences (10,94\%) in a semi-formal context. Finally, there are 2 experiences (3,12\%) where the educational context were not described.

\subsubsection{Mobile device availability}

The experiences carried out with the 44 apps were developed with students' own mobile devices (66,67\%). With 5 apps the authors did not specified which mobile phone was used.

\subsubsection{Compulsory use of the apps}
\begin{figure}
	\centering
		\includegraphics[scale=0.45]{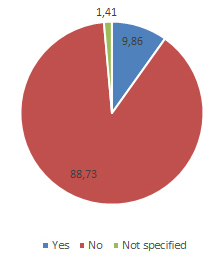}
	\caption{Compulsory use of the apps}
	\label{fig:Compulsory}
\end{figure}

In the different experiences, only the use of 7 apps (9,86\%) was mandatory during the experiences (see Figure~\ref{fig:Compulsory}). In the rest of the experiences, the use of 63 apps (88,73\%) was not mandatory. Finally, we want to mention that there is 1 app where it was not specified whether it use is mandatory or not.

We can also want to remark that only the use of 4 apps was integrated in the curriculum.

\subsubsection{Teacher methodologies}
\begin{figure}
	\centering
		\includegraphics[scale=0.5]{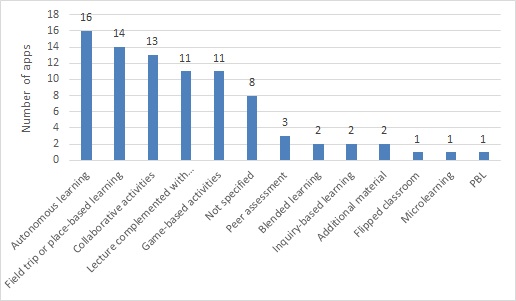}
	\caption{Teacher methodology with the app}
	\label{fig:TeacherMethodology}
\end{figure}

Figure~\ref{fig:TeacherMethodology} shows the methodologies used by teachers. We can see that there are 11 different methodologies. The methodology what was used the most to use with MLW apps is self-directed learning. Next, we can mention teacher-directed learning, collaborative, and game-based learning. We can also point out two interesting facts. First, in 21 experiences more than a methodology was used. Second, in 10 experiences, the authors did not specified the methodology used.

\subsubsection{Learning activities}



We also analysed how long the app was used in the experiences. We have observed that there are apps used in a whole course, apps that were used in single activity of the course or in tests to check some features of the app (some test were developed in real activities and other in test scenarios). We have also seen that there are papers that present more than one evaluation of the app. 

Namely, our analysis show that 16 apps (22,53\%) were used during the whole course, 33 apps that only were tested with students in a real lecture, 20 apps were used only for a single activity in the course with students, 17 apps that were proved in a test (to test if it is useful, usable, etc) with students but not in a real lecture. 

\begin{figure}
	\centering
		\includegraphics[scale=0.40]{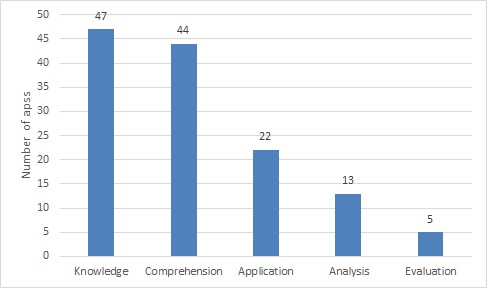}
	\caption{Activities developed by students according Bloom's taxonomy}
	\label{fig:ActivitiesBloom}
\end{figure}

An analysis of the different activities students made with the app classified according to Bloom's Taxonomy is depicted in Figure~\ref{fig:ActivitiesBloom}. As can be seen, in general, the apps covered more than one type of activity and most of the apps were used to work the low levels of Bloom's taxonomy (knowledge and comprehension). Only 5 apps reported the inclusion of evaluation activities.

From 71 apps, 66.2\% apps cover knowledge activities, 66.91\% comprehension, 30.99\% application, 18.31\% analysis, and, finally, only 7.04\% evaluation.
 
\subsubsection{Measured outcomes}

In many experiences, the authors have evaluated whether the apps developed are easy to use for students. In all the apps evaluated (38) as for this issue, students considered that they are easy to use, that is, 100\%. In the rest of the apps (33), an evaluation of the app as for whether it was easy to use or not was not made. Even in some cases, there is a usability analysis of the apps. Namely, in 44 apps an usability analysis was made. The results obtained were: 40 positive (90.91\%), 2 mixed (4.55\%), 1 negative (2.27\%), and 1 neutral (2.27\%).

Students have also been queried about whether they consider that the app helped them to improve their learning. From the 71 apps, in 29 the students were asked about this issue and in all cases (100\%), students answered that the apps improved their learning. In no experience, the global students' assessments indicated that the use of the app did not improve their learning.

There are also experiences that, apart from taking into account students' perception, they made an evaluation of the performance to determine whether the app was really improving students' performance. With 29 apps there was a performance analysis and with 26 apps the results were positive (89,66\%). There also was 1 negative result and 2 neutral/inconclusive results.

The engagement was evaluated through 37 apps. The results show that in 35 cases the evaluation was positive, there is 1 conditional positive and 1 inconclusive result. In rest (34 apps), engagement was not evaluated.

\subsubsection{Outcomes of the results}
\label{sec:outcomesresults}

\begin{figure}
	\centering
		\includegraphics[scale=0.4]{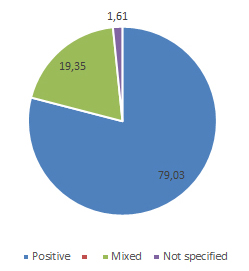}
		\caption{Results of the experiences}
	\label{fig:ResultsExperiences}
\end{figure}

Most of the papers report positive outcomes in the development of m-learning experiences (see Figure~\ref{fig:ResultsExperiences}). Namely, 52 experiences (81,25\%) report positive results over the different issues evaluated. The rest of them report mixed results (12 experiences - 18,75\%). 

Our analysis of experiences with mixed results present different issues we mention next. There are experiences where they found some usability issues~\cite{Abd_Husain_2014,Kim_Kim_Han_2013,Lehmann_Sollner_2014}, that the comparison between students' performance on the mobile and desktop tests are similar~\cite{Bogdanovic_Barac_Jovanic_Popovic_Radenkovic_2014} or results that did not show that the use of the app results in a better understanding~\cite{Browne_Anand_2013}.~\cite{Schmitz_Klemke_Walhout_Specht_2015} also indicate that learning using desktop computers differs from using mobile devices.

\cite{Boticki_Baksa_Seow_Looi_2015} tested the use of badges and they found that badges can only motivate students to learn in a meaningful way when teachers provide a suitable learning context. 

\cite{Cocciolo_Rabina_2013} show that the incorporation augmented reality did not contribute to their understanding or engagement (it seems that is due to usability problems with the app). 

\cite{Huang_Yang_Chiang_Su_2016} presented that there were not an improvement students’ attention to learning materials using mobile phone.~\cite{Martin_Ertzberger_2013} also found in their experience that computer-based instruction (CBI) outperformed the instruction with iPad and iPod and that there was no significant difference between iPod and CBI. 

Other experiences that show inconclusive or not significant results regarding the performance are~\cite{Pereira_2016,Shu-Chun_Sheng-Wen_Pei-Chen_Cheng-Ming_2017}. Finally, in the experience developed by~\cite{Rensing_Tittel_Steinmetz_2012}, they commented that for a task in their experiences students did not see, apparently, the benefit of using a smart phone.

\section{Discussion}
\label{sec:discussion}

In this section we present the results we have obtained once we have applied the SLR procedure described in previous section. Next, we answer the different questions we defined for our study (see Section~\ref{sec:introduction}).

\subsection{RQ1. In which educational context were used the mobile (web) learning apps?}

After analysing the results obtained we can conclude that the experiences were focused on a formal context (see Figure~\ref{fig:EducationalContext}), mainly, in the university (see Figure~\ref{fig:LearnerLevel}), where the activities with MLW apps were made in the classroom or in activities related to the lectures that were made in the classroom (planned outdoor activity or in blended learning) (see Figure~\ref{fig:LearningSetting}). The main disciplines where these activities were mainly made were Engineering and Humanities. Particularly, the main area was computer science. The experiences were developed during a short period of time (less than 10 weeks) (see Figure~\ref{fig:ExperienceDuration}), where an important number of them were based on a single session.

A formal context for the classes is a result that could be expected since more of the scientific publications were based on research developed in universities. This is also shown in previous SLRs. Indeed, in~\cite{Crompton_Burke_Gregory_2017} formal context represented 50\% of the studies, in~\cite{Crompton_Burke_2018} 54\%, and 44\% in~\cite{Bano_Zowghi_Kearney_Schuck_Aubusson_2018} (even it is not higher than 50\% we should take into account that it is the main category in this review and that is also considered in other classifications where several education contexts are used with a 27\%). The only review that report different results comes from~\cite{Chee_Yahaya_Ibrahim_Hasan_2017} where informal learning obtained higher values (11.11\% vs 8.33\%). These results show that more research should be made in informal or semi-formal contexts.

The use of the mobile devices in classroom is justified since they allow for gathering information in real-time, collaboration, mobility within classroom spaces or because they can be useful for flipped classroom~\cite{Baran_2014}.~\cite{Sung_Chang_Liu_2016} present similar results since they indicate that in their review, most of the experiences they analysed were made in the classroom (50.0\%), followed by those that were made outdoors (15.5\%) and, finally, some of them were made in unrestricted settings (16.4\%). In~\cite{Bano_Zowghi_Kearney_Schuck_Aubusson_2018,fu2018trends} is also shown the dominance of m-learning in classroom and laboratories, and~\cite{Bano_Zowghi_Kearney_Schuck_Aubusson_2018} point out that only one study takes advantage of mobility as a key advantage of m-learning. This concern is also shown in~\cite{Zydney_Warner_2016}.~\cite{Lai_2019} also reports the use of m-learning in classroom, although in their survey, the biggest group was learning in different contexts, followed by those that did not conducted learning activities and those made in real context, which are also the main group in~\cite{Chung_Hwang_Lai_2019}. In our case, these categories are also represented with a 29.03\% and 30.64\%, respectively. We can also point out that, unlike previous surveys, a number of experiences were only made in a test. This was due to that in many cases, the experience was developed to test the app since it was a part of the app development process.
 
The use of the mobile devices in classroom is justified since they allow for gathering information in real-time, collaboration, mobility within classroom spaces or because they can be useful for flipped classroom~\cite{Baran_2014}.~\cite{Sung_Chang_Liu_2016} present similar results since they indicate that in their review, most of the experiences they analysed were made in the classroom (50.0\%), followed by those that were made outdoors (15.5\%) and, finally, some of them were made in unrestricted settings (16.4\%). In~\cite{Bano_Zowghi_Kearney_Schuck_Aubusson_2018,fu2018trends} is also shown the dominance of m-learning in classroom and laboratories, and~\cite{Bano_Zowghi_Kearney_Schuck_Aubusson_2018} point out that only one study takes advantage of mobility as a key advantage of m-learning. This concern is also shown in~\cite{Zydney_Warner_2016}.~\cite{Lai_2019} also reports the use of m-learning in classroom, although in their survey, the biggest group was learning in different contexts, followed by those that did not conducted learning activities and those made in real context, which are also the main group in~\cite{Chung_Hwang_Lai_2019}. In our case, these categories are also represented with a 29.03\% and 30.64\%, respectively. We can also point out that, unlike previous surveys, a number of experiences were only made in a test. This was due to that in many cases, the experience was developed to test the app since it was a part of the app development proces. 

Although the highest number of experiences was made within the classroom, it is followed very closed by experiences that were made out of the classroom or in field trips. In this type of experiences, students can take advantage of mobilized learning~\cite{Liu_Scordino_Geurtz_Navarrete_Ko_Lim_2014}: multiple entry points and learning paths supporting differentiated learning, providing students possibilities to create artifacts according their needs, enabling that student to improvise according their need in the learning context, and supporting students can create and share artifacts on the move. Thus, learning can be made outside the classroom and expand their school/university day working at their own pace, monitoring their own progress and accessing to additional resources~\cite{Liu_Scordino_Geurtz_Navarrete_Ko_Lim_2014}. The m-learning which is context-aware and offer adaptive support is named situated learning and it has shown to provide improvement of learning achievement and learning attitude~\cite{Santos_etal_2014}. As pointed by these authors, the problem is how to include this kind of learning within the curriculum since, so far, they have been considered as informal learning activities. They consider that they should be integrated withing the curriculum to increase the impact of these activities in the learning outcomes.

As for domain subject,~\cite{Sung_Chang_Liu_2016} presents that the main domain was language arts, followed by science, mathematics, and computer and information technology. Next, appeared others with quite less relevance. In~\cite{Lai_2019}'s survey, the classification of domain are: science, language, social science, engineering or computers and others. Although our results are different, they are also aligned with them. 

Regarding the duration of the experience, our results are partially aligned with some previous work such as in~\cite{Sung_Chang_Liu_2016}, where the most frequent period was experiences between 1 and 6 months, followed by between 1 and 4 weeks. In our review, the most frequent period was less than 10 weeks, followed by experiences that were made in only one session. There is some difference between periods but both coincide that results based on long-term periods (courses that are made during a whole academic year or several academic years) are not common but, at least, there were experiences that seem to cover a whole semester or a four-month period, representing the last of a whole course. Therefore, longer research studies should be made to have a clear perception on the possible benefits of using mobile apps in the learning process. Similar recommendations have arose in previous works such as~\cite{Zydney_Warner_2016},~\cite{Sung_Chang_Liu_2016} or~\cite{fu2018trends} where the authors indicate that we should take into account the effect that could be produced by the novelty for technology or in~\cite{Hwang_Fu_2019} which indicates, in the application of m-learning for learning language, the need of conducting more research works which involve both larger number of students and longer duration of the experience. Finally,~\cite{Bogdanovic_Barac_Jovanic_Popovic_Radenkovic_2014} concluded that m-learning activities should be frequent, based on a practical content and brief to avoid problems related to battery life, slow input and fatigue. This is an issue that should be reconsidered since the new generation of smartphones has significantly improved features such as battery life and an input.

\subsection{RQ2. What was the purpose of the mobile learning (web) apps developed?}

Our results show that the apps were mainly CLA and that there are apps that cover more than a purpose being CLA and SLA the main combination (see Figures~\ref{fig:TypeAppsPurpose} and~\ref{fig:TypeAppsPurposeSingleClasses}). 

\begin{figure*}
	\centering
		\includegraphics[scale=0.4]{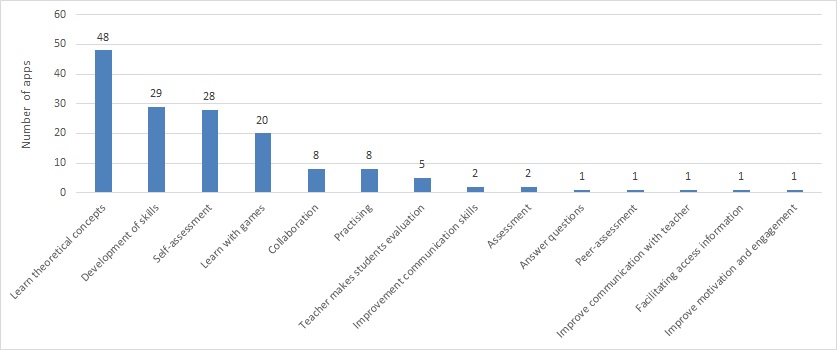}
	\caption{Purposes of the apps}
	\label{fig:PurposesApps}
\end{figure*}

Apart from these data, we have made a more detailed analysis of the different purposes the apps were used based on the authors' description. This analysis is shown in Figure~\ref{fig:PurposesApps}. We can point out that an app can be used for multiple purposes and that most of the apps were leveraged so that student learn theoretical concepts (67.61\%). Furthermore, we can find that apps were leveraged for the development of skills (40.85\%) or for self-assessment (39.44\%) and applications that took advantage of gamification (28.17\%). This results are consistent with the results just mentioned regarding CLA and SLA.
 

Considering the most important results presented so far, in a general way, we could state that the MLW apps developed were created to be used voluntarily by students (see Figure~\ref{fig:Compulsory}) as a new way to learn and increase the engagement (Figure~\ref{fig:FigureOutcomesEvaluated}) by means of the development of CLA in many cases combined with SLA being used in the classroom, out of the classroom, or in field trips. For this reason, these apps incorporated as features (Figure~\ref{fig:AppsFunctions}): images, content, providing some of interactivity, quizzes, audio, and exercises. As explained by~\cite{Zydney_Warner_2016}, multimedia capabilities are used to provide visual and/or audio representations of information. The features of the apps is an issue that has been scarce analysed in the literature.

These results are consistent with the study presented by~\cite{Domingo_Gargante_2016}, which indicates that the group of apps more frequently used is CLA. In this same study, the authors indicate that teachers contemplated m-learning mainly for facilitating access to information followed by new ways to learn, next, an increase in engagement and, finally, collaborative learning. There are similarities regarding the importance of new ways of learning and the increase in engagement. By contrast, in our study facilitating access to information was the least appreciated feature. From our point of view, this is due to that the apps analyzed here were created purposely to learn a specific content or skill and they were not general apps that could be used for learning different issues. As pointed out by~\cite{Chung_Hwang_Lai_2019} there is still a large of research works that leverage technology to provide content.


\subsection{RQ3. What were the results obtained in the experiences with the mobile learning (web) apps (positive, negative, mixed)?}

The data depicted in Figure~\ref{fig:ResultsExperiences} and commented in Section~\ref{sec:outcomesresults} show clearly that the results obtained in the different experiences are positive (79.03\%). The comparison of our results based on m-learning experiences developed with specific developed apps is quite similar to SLRs on m-learning experiences presented in related work. Namely,~\cite{Wu_Jim_Wu_Chen_Kao_Lin_Huang_2012} reports 86\% of positive results,~\cite{Liu_Scordino_Geurtz_Navarrete_Ko_Lim_2014} a 75\%,~\cite{Zydney_Warner_2016} points out that the 87\% of the studies present, at least, one learning outcome as statistically significant, and~\cite{Crompton_Burke_Gregory_2017} report a 70\% of positive results in PK-12 education,~\cite{Crompton_Burke_2018} report a 70\% of positive results in higher education,~\cite{Bano_Zowghi_Kearney_Schuck_Aubusson_2018} a 86\%, and~\cite{Lai_2019} a 75.24\%. Therefore, we can see that our results are within the same range. In these SLRs, they cover both generic and specific apps, and we are seeing that the isolation of specific apps do not make a significant change. In related work they only survey that reports not so clear good results is from~\cite{Chee_Yahaya_Ibrahim_Hasan_2017} that only achieved a 52.6\% of positive outcomes between 2010 and 2015.

We consider that it is interesting to point out that papers with negative results are hardly published. Indeed,~\cite{Wu_Jim_Wu_Chen_Kao_Lin_Huang_2012},~\cite{Crompton_Burke_Gregory_2017} and~\cite{Lai_2019} report only a 1\% of negative outcomes.~\cite{Chee_Yahaya_Ibrahim_Hasan_2017} reports a 3.25\%, in~\cite{Crompton_Burke_2018} they show a 4\%, and in~\cite{Bano_Zowghi_Kearney_Schuck_Aubusson_2018} a 2\%. In our research, we have not found any paper with only negative results.

\subsection{RQ4. What were the learning improvements (if there) that we were achieved with the use of specific mobile (web) applications and with were related to?}

\begin{figure*}
	\centering
		\includegraphics[scale=0.5]{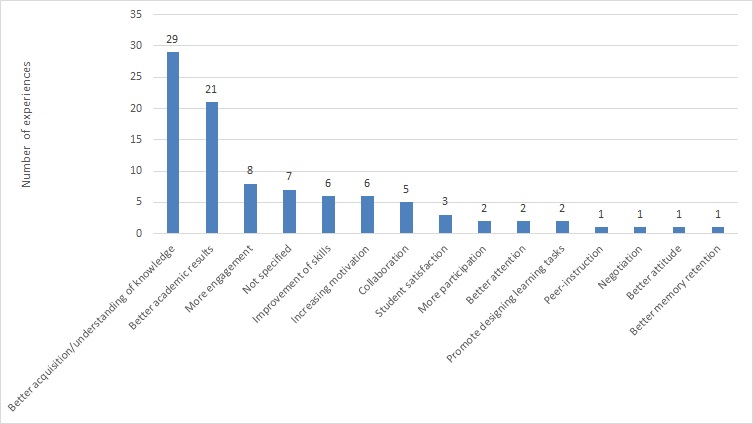}
	\caption{Research outcomes}
	\label{fig:ResearchOutcomes}
\end{figure*}

The analysis of the learning outcomes that the experiences reported are shown in Figure~\ref{fig:ResearchOutcomes}. As can be seen, the MLW apps mainly were used to evaluate whether students achieved a better understanding of knowledge (30.53\%) and a better performance (22.11\%). Next, with a less significant values (ranging from 8.42\% to 1.05\%), they were used to improve attitudes or traversal skills, e.g., engagement, motivation, collaboration, negotiation, etc.

Regarding the understanding of knowledge,~\cite{Browne_Anand_2013} points out that MLW apps should be used to practise the understanding of concepts and should incorporate gamification to increase user satisfaction. Gamification has also shown an increment in the learning process as for: enjoyment, engagement, and having studying students more active~\cite{Su_Cheng_2015}. Performance and attitudes will be analysed below.


\begin{figure*}
	\centering
		\includegraphics[scale=0.55]{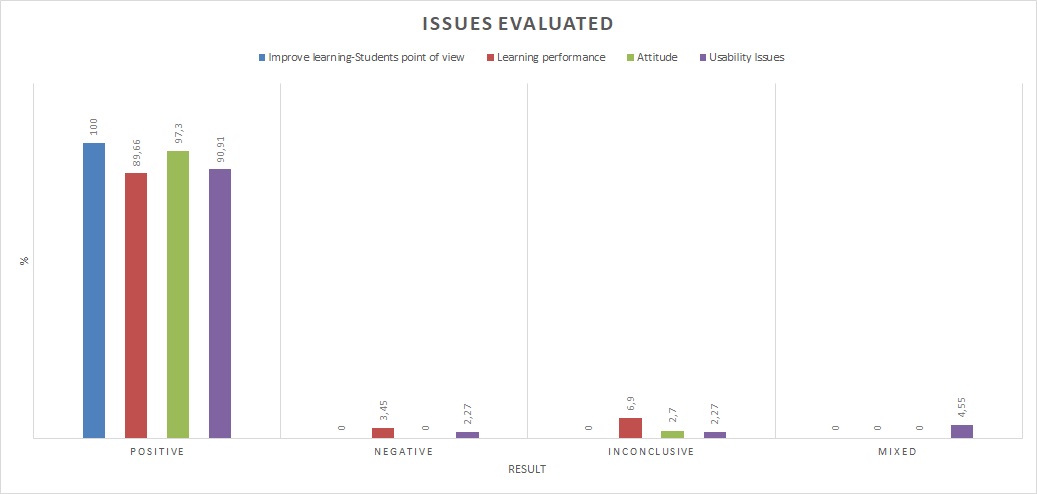}
	\caption{Issues Evaluated}
	\label{fig:RQ-IssuesEvaluated}
\end{figure*}

Figure~\ref{fig:RQ-IssuesEvaluated} shows the results of grouping the use of the apps in the different experiences into four main categories considering students' point of view: whether students considered that their learning was improved with the use of the app (evaluation made in 29 apps), whether they obtained a better learning performance (based on numerical evidence in 29 apps) by means of the app, whether attitude (motivation, engagement, etc) improved using the app (in 37 apps) and, finally, whether apps were easy to use (in 44 apps). 


The analysis shows that the positive results were achieved mainly in usability issues, followed by attitude and, finally, in an equal way in improving learning and learning performance.

It is common that usability tests were made since an important part in the process of software development of an app is usability issues. If the app was not designed bear in mind this issue, the app would not be used and, therefore, it would not be possible take advantage of m-learning. Indeed, as pointed out by~\cite{hsiao2018understanding_080} when users have the believe that a system was able to be used in an open, easy and speedy way, they increase their intention to use it. For this purpose, formal methods for measuring usability were used such as Nielsen's 10 usability heuristics~\cite{nielsen1992finding}, the Questionnaire for User Interaction Satisfaction(QUIS)~\cite{chin1988development}, the ''Five Es'' of usability of a product~\cite{zhang2005challenges}, the System Usability Scale (SUS)~\cite{bangor2008empirical} or the Shneiderman et al.'s guidelines~\cite{shneiderman2010designing} in several works such as~\cite{Dirin_Nieminen_2015,Cabielles-Hernandez_etal_2014,Correa_etal_2013,Kim_Kim_Han_2013,Schmitz_Klemke_Walhout_Specht_2015,Zbick_Nake_Milrad_Jansen_2015,Sun_Chang_2016}.

From our point of view, the drawback is that in many cases, the experiences paid more attention to this issue than teaching-learning issues. 


Regarding attitude (motivation, engagement, ...), 97.3\% of apps used have shown positive results. Only one app stated inconclusive results. Attitude was studied since it was considered that learners would invest more cognitive resources in the learning experience if they were more engaged~\cite{Cocciolo_Rabina_2013}. Motivation is an important issue on learning process and in m-learnign has proved that students can achieve higher scores, meta-cognition, and psychological need satisfaction~\cite{koh2010investigating,Jeno_Grytnes_Vandvik_2017}. Indeed,~\cite{Su_Cheng_2015}'s study shows that there is a positive relationship between learning outcomes and motivation, and~\cite{Jeno_Grytnes_Vandvik_2017} found a significant difference in intrinsic motivation and perceived competence when compared students using an app with students using a textbook.~\cite{Melero_Hernandez-Leo_Manatunga_2015} states that the use of smartphones shown that students were more engaged, less distracted and more focused on the activity. Their results also show that engagement is influenced by the amount of students per group when they are participating in a group-based activity using smartphones.

\cite{hsiao2018understanding_080} results show that positive effect on learning outcomes and motivation can be achieved when a context-aware u-learning environment is combined with ubiquitous games. They also found out that an emotional involvement leads to a significant positive impact on the intention of the usage of the system. Motivation in mobile-based assessment has analysed~\cite{Nikou_Economides_2018} and their findings show that its positive impact is significant. Although they also mentioned that is also needed more research on its influence on learning outcomes.~\cite{Jeno_Grytnes_Vandvik_2017} also points out that we also need to better understand why learners are more intrinsically motivated when mobile applications are used.


Students' perception as for learning improvement is that their learning had improved (in 29 apps that this issue was asked, all the results are positive, i.e. 100\%). Finally, there are also 29 apps that obtained data to measure performance. The results indicate that in 26 cases the results were positive (89.66\%). The rest of the experiences with the apps show: 1 negative experience, 1 neutral, and 1 inconclusive. The negative results were reported in~\cite{Martin_Ertzberger_2013}, where authors indicate that students following a computer-based instruction (CBI) outperformed users using iPad/iPod. They note that students in CBI were less distracted and that the novelty of using iPad/iPod could be the cause so that these users obtained less scores. Other possible reasons they noted are how the information is processed and some conditions related to context.


It is important to point out that, although learning performance was not proven to be better, the results show that students perceived that their learning had improved, and they were more engaged, excited and motivated~\cite{Martin_Ertzberger_2013}.

Obtaining a better performance with mobile devices is an expected result since a detailed studied developed by~\cite{Sung_Chang_Liu_2016} revealed that the effect of mobile devices in performance is better than desktop computers or not using mobile devices as intervention. They also pointed out that its use is more effective in inquiry-oriented learning than its usage along lectures, self-directed study, cooperative learning, and game-based learning.~\cite{Jeno_Grytnes_Vandvik_2017} also compared the performance between learning with an app and a textbook and the results shown a significant high score in favour of those that used an app. Its leverage in group-based activities has also shown a significant positive impact on the performance students achieve in the activity.~\cite{Nikou_Economides_2018} presented in a study on mobile-based assessment that only found 1 negative result and 2 neutral results. The rest of the cases where there was an evaluation of learning performance were positive (86.65\%). Their results almost identical to ours.

Regarding performance it is also importat to point out that in some cases to have better results could be due that students had passed more time with the app (e.g. due to include gamification features)~\cite{Palomo-Duarte_Berns_Dodero_Cejas_2014}.

Another issue that is dealt with in the papers analysed but that has also reported positive results, as collaboration, has also been analysed into detail by~\cite{fu2018trends} and these authors show that mobile-based collaborative learning provide better results than Internet-based learning and that can help students for better understanding and conception transformation. Indeed, they state that is a research field that is rapidly growing.

Our results are aligned with results previously obtained in the previous SLR. Indeed,~\cite{Wu_Jim_Wu_Chen_Kao_Lin_Huang_2012} found a 86\% of positive outcomes and from 164 studies, being only 4\% neutral and 1\% negative.~\cite{Liu_Scordino_Geurtz_Navarrete_Ko_Lim_2014} results show that in the comparison between the use or not of mobile devices, 69.25\% of works (9) had better positive outcomes with mobile devices regarding those that followed a traditional instruction. This result is also supported by the survey made by~\cite{Sung_Chang_Liu_2016} that shows better results using mobile devices than desktop computers or not using mobile devices.~\cite{Zydney_Warner_2016} found in 87\% of the studies they reviewed, at least, one learning outcome was statistically significant.

\subsection{RQ5. What pedagogical approaches were followed in the experiences with the apps?}

\begin{figure*}
	\centering
		\includegraphics[scale=0.6]{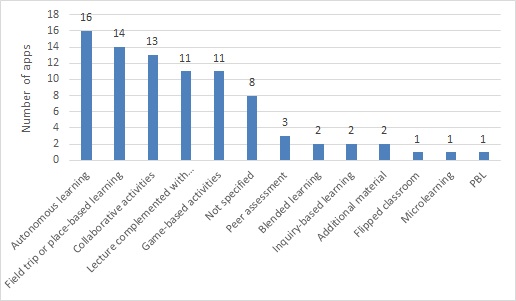}
	\caption{Learning design}
	\label{fig:RQ-LearningDesign}
\end{figure*}

The analysis made on the pedagogical approach followed to develop the learning experience with the apps shows in Figure~\ref{fig:RQ-LearningDesign} that they main use of m-learning apps was to support that students could learn in autonomous way. As pointed by~\cite{subramanya2012point}, apps could be a powerful supplement to traditional teaching and learning. Furthermore, it can extend the learning beyond classroom walls and offer new possibilities to perform authentic learning~\cite{Liu_Scordino_Geurtz_Navarrete_Ko_Lim_2014}. These apps were also used for developing field trips or place-based learning or situated learning, collaborative activities or game-based activities. As pointed in~\cite{Chang_Chatterjea_etal_2012}, learning in a authentic real-context helps learners' understanding.

But, at the same time, there is an important number of apps that were used in the classroom to develop activities based on its usage. This is due to, as mentioned in~\cite{Baran_2014}, it facilitates the movement of students within the classroom physical space and helps to establish flipped classrooms or problem-based learning~\cite{Jou_Lin_Tsai_2016}.~\cite{park2012university}'s study as a conclusion indicates that the incorporation of a MLW app could increase motivation and make students' learning process more convenient and enjoyable than traditional methods.~\cite{Browne_Anand_2013} consider that MLW apps should be used to augment and enhance traditional instruction. As can be seen in Figure~\ref{fig:RQ-LearningDesign}, in our survey, there are apps that were used in the lectures. However, these apps were scarcely leveraged for flipped classroom or problem-based learning. In this study, within the classroom, MLW apps were mainly used for collaborative work.

Other approached followed was the Here and now approach introduced by~\cite{Martin_Ertzberger_2013} where learners could learn anytime at anyplace being this process situated in the context of their learning by means a mobile device. This approach evidenced significant achievement and attitude when compared to computer-based Instruction.

\cite{Bogdanovic_Barac_Jovanic_Popovic_Radenkovic_2014} findings using low-cost devices show that the content used in m-learning should be mainly practical, not time consuming and divided into small chunks of information. This facilitates that content elements can be combined to create sensible units~\cite{Boticki_Barisic_Martin_Drljevic_2013}.~\cite{Browne_Anand_2013} also reached similar results showing that apps are better to practice or reinforce concepts. Their findings show that students consider equally interesting using traditional methods to teach but preference in practical issues for the apps. Furthermore, they indicate that game elements should be incorporated to increase user satisfaction. According to~\cite{Chua_Balkunje_2012} for the design of effective m-learning games it is required blending motivation, gaming and learning features. 

Previous SLRs also point out the introduction of games can facilitate students' engagement to achieve a meaningfully learning of relevant content~\cite{Liu_Scordino_Geurtz_Navarrete_Ko_Lim_2014}.


In our analysis we have also seen that the apps developed were used so that students develop mainly knowledge and comprehension in the Bloom's taxonomy (see Figure~\ref{fig:ActivitiesBloom}). However, few apps and activities with them were developed to improve higher level outcomes.~\cite{Lai_Hwang_2015} found out that by means an interactive peer-assessment criteria development approach is possible to engage students in higher order thinking.~\cite{Zydney_Warner_2016} also pointed out in their survey that much effort has been put in measuring lower-level outcomes and that emphasize should be put into higher-level outcomes. They indicate that this lack of research on this higher-level outcomes could due to it is needed more time-intensive methods and limitations of the research settings. This issue is also pointed out by~\cite{fu2018trends} regarding mobile collaborative learning activities or in mobile language learning~\cite{Hwang_Fu_2019}.~\cite{Sung_Chang_Liu_2016} also agrees in the need of more elaborated instructional design developments to take advantage of possible educational benefits by using mobile devices. Thus, we will realize whether the positive results obtained in the experiences are due to the design made based on m-learning or to the novelty when mobile devices are introduced.


During this learning process, cognitive load should be taken into account and~\cite{Jou_Lin_Tsai_2016} states that m-learning could used to prevent unnecessary increment in it.


From out point of view, generally speaking, pedagogical issues were not considered in a comprehensive way and more studies should be made to explore the application of different instructional designs using MLW apps in order to determine different approaches, to confirm whether active methodologies can take advantage of m-learning both in the classroom and out of it, and to improve the procedures when leveraging them. Furthermore, it is fundamental that in order to determine these effects, longer experiences should be made. This is an issue already mentioned in~\cite{Sung_Chang_Liu_2016}.

\subsection{RQ6. What were the main features that the mobile learning (web) apps developed?}

After the analysis of the apps features made in Section~\ref{sec:appsfeatures}, we can indicate that, in general, the apps were developed for Android devices as CLA. Although there were a number of applications that were not designed for a single purpose and they also covered others. The most common combination was CLA with SLA. These apps mainly incorporated images, content, interactivity, quizzes, audio and exercises. There were also other interesting features we mention below.

These features are aligned with studies that identified the elements required for an app, in this case, for learning Math/Engineering concepts~\cite{subramanya2012point}. Indeed, users have shown preference on learning combining text and images as well as being willing to use interactive functions~\cite{Sun_Chang_2016}.

~\cite{Kim_Kim_Han_2013} designed a widget where they only showed between five and nine contents on a page taking into that short-term memory can only hod this number of items of information.

~\cite{Melero_Hernandez-Leo_Sun_Santos_Blat_2015} states that an interesting feature is the use of learning analytics visualizations since it allows the students to make a better diagnosis of their performance and to the teachers a better enactment of the m-learning activity.

Furthermore,~\cite{Fernandez-Lopez_etal_2013} indicate that personalization of activities is important to satisfy special educational needs.

A feature that progressively is being incorporated is the use of Augmented Reality~\cite{Abd_Husain_2014,Correa_etal_2013} since it enhances users' perception, spatial skills, how they interact with the real world and the scaffolding in learning~\cite{Bacca_Baldiris_Fabregat_Graf_Kinshuk_2014}.

3D in apps was also considered and it proved that was more effective than classical teaching methodology~\cite{Noguera_etal_2013}.

The provision of automated assessment was also considered an important feature to provide immediate feedback~\cite{Ortiz_etal_2015}. In some cases, this assessment and the provision of feedback was made by using quizzes~\cite{Pereira_2016}. The use of MCQ applied in a spaced repetition produce longer memory retention~\cite{Pham_Chen_Nguyen_Hwang_2016}.

The use of location (GPS or indoor) was also used since there were experiences based on field trips, museums, etc. Thus, the learning process is contextualized or situated and there is an increase in the authenticity of learning process~\cite{Rensing_Tittel_Steinmetz_2012,Santos_etal_2014}.

Sharing information and collaboration was also considered interesting since it promotes active learning~\cite{Rensing_Tittel_Steinmetz_2012}. Furthermore, when conceptual cooperative learning strategies are applied, students can learn in a more flexible and interactive way~\cite{fu2018trends}.

In geolocated questions/tasks the incorporation of hints or descriptions proved to be useful for the students to find physical objects that had to be explored.

It is also important the design and the layout since it facilitates the understanding of the information and can get rid of confusion and frustration~\cite{Skiada_Soroniati_Gardeli_Zissis_2014}. Other issues that should also be taken into account are~\cite{Sun_Chang_2016}: font sizes, simple interface and sources with gool credibility and reputation. 

The inclusion of annotations was a feature that was considered as a learning strategy that is useful so that students can review and improve their comprehension of the learning content~\cite{Sung_Hwang_Liu_Chiu_2014} and it is a convenient option when combined with microlecture video~\cite{Wen_Zhang_2015}.

As commented by~\cite{Wang_2013}, including mobile navigation in learning activities can increase learning motivation, interest and interaction. They also remark that learning motivation is positively influenced by joyful learning.

It is important to remark that these features by itself are not sufficient conditions to enhance pedagogies~\cite{Shu-Chun_Sheng-Wen_Pei-Chen_Cheng-Ming_2017}.

This study shows that in the design of apps, new advanced features are being incorporated to offer new possibilities in the teaching-learning process such as virtual reality, augmented reality, or TTS. This shift in the use of the latest capabilities was initially mentioned by~\cite{Zydney_Warner_2016}. This is due to mobile devices currently are powerful devices where we can do almost the same things that desktop computers and that incorporate additional features such as GPS, accelerometers, etc that can make apps take advantage of mobility features.

\subsection{RQ7. What were the development environments used to create these apps?}

As presented in Section~\ref{sec:appsdevelopmenttools}, in general, MLW apps were developed using specific tools for the MOS for which the apps was developed. This kind of development has several benefits: it improves user's experience since the user interface generated is similar to the rest of the apps of the MOS and, in general, the performance is better. The main disadvantage of developing an app for an specific MOS is that to cover all the different students' MOS requires the development of different apps using different tools. This is costly and makes difficult the maintenance of the app. Indeed, a multiplatform design is recommended for mobile persuasive applications that have been developed to develop social and life skills in autistic children~\cite{Ahmad_Shahid_2015}.  

Some mobile educational apps were based on the model client-server since, according to~\cite{parsons2007software}, it is the most flexible and popular solution for these kind of apps. They also indicate that the development of the apps should be done for the different platforms although it has some disadvantages as the duplication of code. This could be addressed by using cross-platform mobile development tools. Others have decided to use Web applications since they are independent of operating system~\cite{Wald_Li_Draffan_2014,Ortiz_etal_2015,Wang_2016}.

~\cite{Kim_Kim_Han_2013} instead of developing an app they decided to develop a widget since it could be used in a mobile device, in a website or in a desktop computer. As these authors pointed out, when the widget is merged with the information-processing model, the widget is able to support individual learning, mobile learning, and user-centered learning.

Other interesting approach was the development of a bot for Telegram. Namely,~\cite{Pereira_2016} developed a quizzing bot. They argue that instead of using LMS for everything we should have simpler tools. On the other hand,~\cite{Boticki_Baksa_Seow_Looi_2015} point out that mobile technologies should be integrated in LMS in an unobstrusive way that stimulates students and engaged them on repeated use. In our study, we have seen that in very few cases this integration has been produced and apps are used independently.


Finally, there are apps that were developed using authoring tools since they did not require teachers to have a technical background to design and developed their own mobile applications~\cite{Zbick_Nake_Milrad_Jansen_2015}.~\cite{Santos_etal_2014} use a extension of QuestTInSitu to develop a m-learning activity for supporting formative assessment in situ.~\cite{Vasquez-Ramirez_etal_2014} developed AthenaTV for generating educational applications for TV based on the Android OS.~\cite{Schmitz_Klemke_Walhout_Specht_2015} use ARLearn, which supports the design of mobile games that are based on interactive location.~\cite{Wang_2016} uses the commercial tool Adobe Captivate.

For the design and development of the apps, in many cases prototypes and iterative cycles of design, evaluation and redesign were used~\cite{Rubegni_Landoni_2014}. As stated by~\cite{Santos_etal_2014} how we leverage technologies has a direct influence in the design of the activity.~\cite{subramanya2012point} also point out the importance of the design of the app so that it does not add a burden to the learner. Indeed, usability has been considered an important issue and several methodologies have been used as we have previously stated: Nielsen's 10 usability heuristics, QUIS, the ''Five Es'' of usability of a product, SUS or the Shneiderman et al.'s guidelines~\cite{shneiderman2010designing}.

\section{Conclusions}
\label{sec:conclusions}

Our daily lives cannot be understood without the use of mobile devices. They have supposed a revolution in many fields and COVID-19 will do that their usage is more common. However, in education, this process is evolving slowly. First, through general apps and now steadily by MLW apps, we are seeing that m-learning is coming to be part of teaching and learning processes. As part of this process, we have observed that m-learning more and more is being conducted through specific apps designed for learning (MLW apps) for this reason we decided to characterize the first apps that followed this approach where teachers develop them (the early adopters). So far there was no study on the development of MLW apps, its purpose, features, and methodologies used.

By means of a SLR covering 62 papers and 71 apps developed by early adopters, we have seen that these apps were developed to be used both out of the classroom to develop autonomous learning or field trips, and in the classroom, mainly, for collaborative activities. Then, in spite of the fact that we could initially consider that mobile devices in m-learning has more sense out of the classroom, we have seen that, in the classroom, they also offer interesting possibilities that also increase engagement. In most of the cases, these MLW apps were introduced as a new way to learn. Even though, from our point of view, more studies are required to explore the possibilities of m-learning out of the classroom delving into instructional designs that develop skills and take into account features such as virtual reality, augmented reality or location-aware activities. We have seen that these m-learning apps were not taking advantage of mobility features in all its potential to develop learning activities that were made anytime at anyplace and taking into account context and realistic situations. Furthermore, more apps and instructional designs should be made to develop higher level outcomes instead of low level ones. We have seen that most of the apps were only focused on knowledge and/or comprehension. We want to remark that the results obtained with specific developed apps are quite similar to previous general surveys that are not focused on this type of apps.

We have also observed that, with different instructional designs, m-learning can produce positive results improving learning, learning performance, and attitude. As for learning performance, our findings show that most of experiences report positive results (75-85\%) and there are some negative or inconclusive results but students' perception is that m-learning improves their learning. We have not found any experience showing perception of no improvement. However, we should take into account that these results regarding improvement of learning, learning performance and attitude have obtained based mainly on short m-learning experiences. Then, these positive results could be due to the novelty in using mobile devices or because students expend more time with the app due to a more engagement. Therefore, it is important to develop long term experiences that measure, first, these issues when technology is not already a novelty and, second, the different effects that are produced in these long experiences. Moreover, these m-learning instructional designs should be integrated in the curriculum and developed with different type of learners since most of experiences developed with MLW apps have been developed for university students. Experiences considering its application in lifelong learning should also be addressed since it seems to be a suitable way to develop learning process that can be adapted to learners and that can be developed anywhere at anytime.

These apps were mainly developed using native environments and for specific MOS that require some expertise in programming mobile apps. In the past few years, some frameworks that do not require teachers to have a high level in programming have emerged to make easy teachers the development of these apps. The improvement and widespread of these frameworks would allow a wider leverage of mobile learning. These framework should support to include content, multimedia elements, interactive elements, social elements, different elements for assessments, and advances features that soon will be incorporated usually such as virtual reality, augmented reality, annotations or text to speech or speech to text. A study on the role of these frameworks in facilitating teachers the development of apps that adapts their needs should be addressed.

Finally, we consider that a meta-analysis on systematic literature reviews should be made to analyse all the different issues that are analysed and, in base of these information, to define a comparison framework that indicates the different elements that should be taken into account in a survey on m-learning and that will allow to compare results (e.g. data to be gathered for each experience, what a short experience is considered, classification of the results, classification of the apps, etc).

\section*{Acknowledgment}

The authors would like to thank to University of Murcia for its support to the teaching innovation group no. 40.

\ifCLASSOPTIONcaptionsoff
  \newpage
\fi



%

\bibliographystyle{IEEEtran}
\bibliography{article}

\begin{IEEEbiography}[{\includegraphics[width=1in,height=1.25in,clip,keepaspectratio]{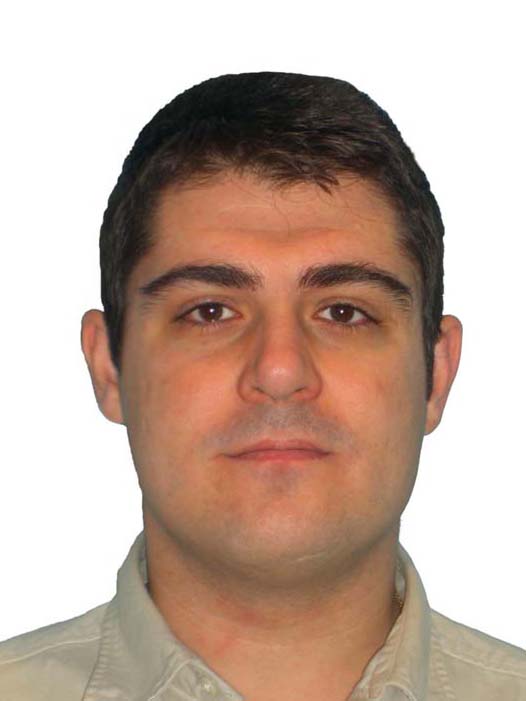}}]{Antonio Ruiz-Martínez}
is an associate professor in the Department of Information and Communications Engineering at the University of Murcia (Spain). He received B.E., M.E. and Ph.D. degrees in Computer Sciences from the University of Murcia. He is also an IEEE senior member. His main research interests include electronic payment systems, security, privacy, and educational technology. He has participated in several research projects in the national and international areas such as ECOSPACE, SEMIRAMIS, INTER-TRUST, and STORK 2.0. He has published more than 50 papers in conferences and journals. He is also serving as an active technical program committee member in 12 conferences (ICC CISS, ICACCI, SpaCCS, SSCC, etc) and, reviewer and member of the editorial board in several international journals. He is leading the teaching innovation group of the UMU in the teaching of ICT and their fundamentals.
\end{IEEEbiography}

\begin{IEEEbiography}[{\includegraphics[width=1in,height=1.25in,clip,keepaspectratio]{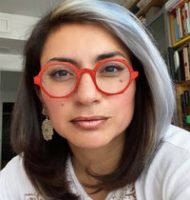}}]{Linda Castañeda}
holds a Ph.D. in educational technology and is an Associate Professor in educational technology at the Faculty of Education of the University of Murcia, in Spain. Thanks to her educational background in education, she has a strong interest in making EdTech research more educational. A participant in national and international research projects on the implementation and impact of technology in both formal and non-formal learning contexts, Linda has closely collaborated with different research institutions in Europe and abroad, and continues to work with institutional initiatives as an advisor. Dr. Castañeda has been a Visiting Scholar at KMi at OU, GSE of UCBerkeley, among others. She is an editorial board member of various international academic journals, has published papers and book chapters in both Spanish and English, and is a member of the Association EDUTEC, NOVADORS, and the PLEConf Community. Linda’s current research portfolio includes critical perspectives on educational technology, competencies for the digital era (definition, development and assessment), strategical approaches to teachers’ professional development, socio-material perspectives of emergent pedagogies, and Personal Learning Environments.
\end{IEEEbiography}


\begin{IEEEbiography}[{\includegraphics[width=1in,height=1.25in,clip,keepaspectratio]{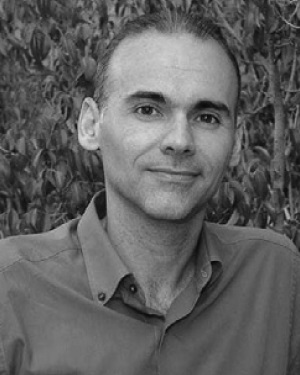}}]{Jesualdo T. Fernández Breis}
(Member, IEEE) received the Computer Engineering and Ph.D. degrees in computer science from the University of Murcia, Spain, in 1999 and 2003, respectively. He is currently a Full Professor with the Faculty of Computer Science, University of Murcia. He is also a member of the IMIB-Arrixaca Bio-Health Research Institute. He has been leading research projects related to semantic web technologies since 2004. His current research interests include the application of semantic technologies for the development of learning health systems and the development of quality assurance methods for ontologies and terminologies.
\end{IEEEbiography}




\end{document}